\documentclass[english]{article}
\usepackage[T1]{fontenc}
\usepackage[latin9]{inputenc}
\usepackage{amsmath}
\usepackage{amssymb}

\makeatletter
\newcommand{\lyxaddress}[1]{
\par {\raggedright #1
\vspace{1.4em}
\noindent\par}
}

\@ifundefined{definecolor}
 {\@ifundefined{definecolor}
 {\@ifundefined{definecolor}
 {\@ifundefined{definecolor}
 {\@ifundefined{definecolor}
 {\usepackage{color}}{}
}{}
}{}
}{}
}{}

\newcommand{\beqs}{\begin{eqnarray}}
\newcommand{\eeqs}{\nonumber\end{eqnarray}}
\newcommand{\beq}{\begin{eqnarray}}
\newcommand{\eeq}{\end{eqnarray}}

\def\ben{\begin{enumerate}}
\def\een{\end{enumerate}}

\makeatother

\usepackage{babel}

\begin{document}

\title{Geometry of the Motion of Ideal Fluids and Rigid Bodies}

\author{S. G. Rajeev%
\thanks{rajeev@pas.rochester.edu%
} }

\maketitle

\lyxaddress{Department of Physics and Astronomy}
\lyxaddress{Department of Mathematics}
\lyxaddress{University of Rochester, Rochester NY 14618}
\begin{abstract}
Arnold pointed out that the Euler equation of incompressible ideal
hydrodynamics describes geodesics on the group of volume-preserving
diffeomorphisms. A simple analogue is the Euler equation for a rigid
body, which is the geodesic equation on the rotation group with respect
to a metric determined by the moment of inertia. The metric on the
group is left-invariant but not right-invariant. We will reduce the
geometry of such groups (using techniques popularized by Milnor) to
algebra on their tangent space. In particular, the curvature can be
expressed as a biquadratic form on the Lie algebra. Arnold's result
that motion of incompressible fluids has instabilities (due to the
sectional curvature being negative) can be recovered more simply.
Surprisingly, such an instability arises in rigid body mechanics as
well: the metric on SO(3) corresponding to the moment of inertia
of a thin cylinder (coin) has negative sectional curvature in one
tangent plane.

Both ideal fluids and rigid bodies can be thought of as hamiltonian
systems with a quadratic hamiltonian, but whose Poisson brackets are
those of a non-nilpotent Lie algebra. We will also describe a different
point of view towards three dimensional incompressible flow in terms
of the Clebsch parametrization. In this picture, the Poisson brackets
are represented canonically. The hamiltonian is represented by
a quartic function.

This is meant mainly  as an expository article, aimed at a mathematical audience
familiar with physics. Based on Lectures at the Chennai Mathematical Institute and
the University of Connecticut.
\end{abstract}
\pagebreak{}

\section{Introduction}

Unlike in mathematics, the problems of physics tend not to be very
old. Experimental advances constantly invalidate old ideas or introduce
completely new ones. However, the problem of understanding a non-integrable
dynamical system (`chaos') is as old as physics itself and is still
largely unsolved. A particularly virulent example is the phenomenon
of turbulence in fluid mechanics: when velocity exceeds a critical
value, the flow suddenly becomes irregular and unpredictable except
by very fine numerical methods. A theoretical understanding of this
phenomenon, perhaps along the lines of Wilson's theory of second order
phase transitions, remains a great challenge.

A separate question is whether the partial differential equations
of hydrodynamics (Navier-Stokes or Euler) have unique solutions and
how regular they are. This has been recognized as a mathematical challenge
worthy of the best analysts. Even though the two problems in physics
and mathematics are different, one hopes that ideas from one will
cross-fertilize the other.

We begin by reviewing the basic equations of the subject. For simplicity
and brevity, it is hard to beat the classic text by Landau and Lifshitz
\cite{LandauLifshitz}. For the geometrical formulation, the references
are the books by Arnold and Khesin \cite{Arnold} and Kambe \cite{Kambe}
. The expository article by Milnor \cite{Milnor} shows how the geometry
of a group can be reduced to algebraic questions on its tangent space.
This will be important for us because it allows us to avoid defining
an infinite dimensional manifold. We can reduce everything to vector
spaces and linear operators on them: much simpler technically, while
the geometry of manifolds continues to provide powerful intuition
in the infinite dimensional case.

\subsection{Ideal Fluids}

We will consider only ideal non-relativistic fluids; that is, a fluids
whose velocities everywhere are small compared to the velocity of
light and in which losses due to friction (viscosity) are small enough
to be ignored.

\subsubsection{The Two Time Derivatives}

There are two ways of thinking about the time dependence of any physical
quantity in a fluid: at a fixed location in space or, along the flow
of the fluid. The first is the partial derivative $\frac{\partial}{\partial t}$
and the second is the total (or material) derivative $\frac{d}{dt}$.
They are related by

\[
\frac{d}{dt}=\frac{\partial}{\partial t}+\mathcal{L}_{v}\]

where $v$ is the velocity of the fluid and $\mathcal{L}_{v}$ is
the Lie derivative. On a scalar field it is just

\[
\frac{d\phi}{dt}=\frac{\partial\phi}{\partial t}+v^{i}\partial_{i}\phi\]

We will also be interested in the case of a density. A density on
a manifold is simply a differential form of the highest possible rank
( i.e., equal to the dimension of the manifold), $\rho dx^{1}\wedge\cdots dx^{n}$.
Although it has one independent component like a scalar, the Lie derivative
of a density is different from that of a scalar:

\[
\mathcal{L}_{v}\rho=\partial_{i}[\rho v^{i}].\]
 The divergence of a vector field can be defined as

\[
\mathrm{div\ }v=\frac{1}{\rho}\mathcal{L}_{v}\rho=\frac{1}{\rho}\partial_{i}[\rho v^{i}].\]

The time derivative of the velocity field itself requires a different
idea. Since $\mathcal{L}_{v}v=[v,v]=0,$ the Lie derivative does not
capture its variation due to the motion of the fluid. Using a Riemannian
metric $g$ of the manifold in which the fluid is moving ( more precisely,
using its Levi-Civita connection ${\nabla}$) we can find the acceleration
of a fluid element as $\frac{dv}{dt}=\frac{\partial v}{\partial t}+{\nabla}_{v}v$

\[
\frac{dv^{i}}{dt}=\frac{\partial v^{i}}{\partial t}+v^{j}{\nabla}_{i}v^{i}\]

where, as usual,

\[
{\nabla}_{i}v^{j}=\partial_{i}v^{j}+\Gamma_{ik}^{j}v^{k},\quad\Gamma_{ik}^{j}=\frac{1}{2}g^{jl}\left[\partial_{i}g_{lk}+\partial_{k}g_{il}-\partial_{l}g_{ik}\right].\]

\subsubsection{Conservation of Mass}

The first law of motion of the fluid is just the conservation of mass:

\begin{equation}
\frac{\partial\rho}{\partial t}+\frac{\partial}{\partial x^{i}}\left(\rho v^{i}\right)=0\label{eq:ConsMass}\end{equation}

We can write this also as

\[
\frac{d\rho}{dt}=\frac{\partial\rho}{\partial t}+\mathcal{L}_{v}\rho=0\]

in terms of the material derivative. It is clear that this equation
does not make use of the Riemannian metric of the manifold $M$ in
which the fluid is moving.

\subsubsection{Conservation of Momentum}

For a fluid without external forces, Newton's second law gives

\[
\rho\frac{dv_{i}}{dt}=-\nabla_{i}p\]

where $p$ is the pressure and $\frac{d}{dt}$ is the derivative taken
in the co-moving reference frame of the fluid. It includes the explicit
time derivative as well as the change due to the motion of the fluid
element:

\begin{equation}
\frac{\partial v_{i}}{\partial t}+v^{k}{\nabla}_{k}v_{i}=-\frac{1}{\rho}\nabla_{i}p.\label{eq:Euler}\end{equation}

It can be re-expressed as the conservation of momentum density:

\begin{equation}
\frac{\partial(\rho v_{i})}{\partial t}+{\nabla}_{k}T_{i}^{k}=0\label{eq:EulerStressTensor}\end{equation}

where the stress tensor density is

\[
T_{i}^{k}=p\delta_{i}^{k}+\rho v^{k}v_{i}.\]

This equation does use the metric of the underlying space, which is
often take to be Euclidean in physical applications. But there are
physically interesting cases of fluid motion on a curved geometry
as well: the ocean or the atmosphere of a planet can often be thought
of as a fluid on the surface of a sphere.

\subsubsection{Equation of State}

So far we have one scalar equation and one vector equation for the
unknown quantities $\rho,p,v^{i}.$We need one more scalar equation
to have enough information to determine them. This is given by an
equation of state: a relation between pressure and density. A model
that works well in many situations is the law $p=A\rho^{\gamma},$
(polytrope) for some constants $A,\gamma$ characterestic of the fluid
. For the atmosphere in the adiabatic approximation, $\gamma\approx1.4$.

\subsubsection{The Wave Equation of Sound}

Before studying any non-linear equation in depth, we must understand
its linear approximation. If the gradient of the velocity is small
and the departure $\rho_{1}$ of the density from some average value
$\rho_{0}$ is small, the above equations linearize to

\[
\frac{\partial\rho_{1}}{\partial t}+\partial_{i}\left[\rho_{0}v^{i}\right]=0\]

\[
\frac{\partial(\rho_{0}v^{i})}{\partial t}+\kappa{\nabla}^{i}\rho_{1}=0.\]

where $\kappa=\left[\frac{\partial p}{\partial\rho}\right]_{\rho=\rho_{0}}$
. By differentiating the first equation w.r.t. time and putting in
the second we get

\[
\frac{\partial^{2}\rho_{1}}{\partial t^{2}}-\kappa{\nabla}_{i}{\nabla}^{i}\rho_{1}=0\]
 which is the equation for some wave propagating with velocity $c=\sqrt{\kappa}.$These
are sound waves. Thus sound is the infinitesimal manifestation of
fluid flow.

\subsubsection{Isentropic Flows}

If there is a function $w$ (`enthalpy') such that \[
\partial_{i}w=\frac{1}{\rho}\partial_{i}p\]
 the flow is said to be isentropic. It basically means that no heat
is lost or gained by the system. Then the second equation of motion
can be written as

\begin{equation}
\frac{\partial v_{i}}{\partial t}+v^{k}{\nabla}_{k}v_{i}+{\nabla}^{i}w=0.\label{eq:isentropic}\end{equation}

We still need an equation of state giving $w$ as a function of $\rho$.

\subsection{Incompressible Fluids}

If the time dependence of the density is small enough to be ignored,
\[
\frac{\partial\rho}{\partial t}=0\]
 we say that it is \emph{incompressible}. Then compressiblily $\kappa$
tends to infinity: the speed of sound is infinite. More precisely,
the speed of the fluid flow is small compared to the speed of sound.
In this case we get the equations of Euler:

\begin{equation}
\partial_{k}\left[\rho v^{k}\right]=0\label{eq:incompressibilty}\end{equation}

\[
\frac{\partial v^{i}}{\partial t}+v^{j}{\nabla}_{j}v^{i}+\frac{{\nabla}^{i}p}{\rho}=0.\]

The density and the metric of the manifold containing the fluid should
be given; $p$ is a Lagrange multiplier enforcing the condition of
incompressibility \ref{eq:incompressibilty}.

\subsubsection*{Incompressible does not mean constant density}

It is often stated incorrectly that an incompressible fluid must have
constant density. Incompressibility only means that the density is
given as a function of space and is independent of time: it is not
a dynamical variable. The atmosphere of the Earth, for example, has
density decreasing with height; yet the atmospheric flows (winds)
have velocities much lower than the speed of sound. The atmosphere
is an incompressible fluid.

\subsubsection{Boundary Conditions}

We have not said much about the underlying manifold $M$ on which
the fluid flows. In most cases of interest it is a domain of Euclidean
space. But it is useful to consider the more general case of a Riemannian
manifold, as the mathematical concepts are more clear then. Even in
physics, occasionally one is interested in the flow of a fluid on
a curved manifold such as the sphere (ocean currents). In general
the manifold $M$ will have a boundary. For an ideal fluid, it is
sufficient that the velocity field be tangential to the boundary.

In the real world, dissipation (viscosity) cannot be ignored near
the boundary, even if it is small away from the boundary. Thus, the
physically correct boundary condition is that the velocity field must
vanish at the boundary, not just its tangential component. We will
assume this stronger condition in what follows.

Thus, all vector fields will vanish at the boundary. All diffeomorphisms
reduce to the identity at the boundary.

\subsubsection{Two Dimensional Incompressible Fluids}

A particularly interesting example is the case of incompressible fluids
in two dimensions. Geophysical applications include ocean currents
and atmospheric flows which are approximately two dimensional: the
atmosphere and the oceans have a depth small compared to the diameter
of the Earth. As an example we consider the case of flow in the plane
and with $\rho=1:$ a part of the sphere that is small enough that
the curvature can be ignored. We will return to a more general theory
later.

We can eliminate $w$ by taking the curl of the above equation. In
two dimensions, the curl of velocity (`vorticity') is a scalar\[
\omega=\partial_{1}v_{2}-\partial_{2}v_{1}.\]

Since\begin{eqnarray}
\partial_{1}(v_{j}\partial_{j}v_{2})-\partial_{2}\left(v_{j}\partial_{j}v_{1}\right) & = & \partial_{1}v_{j}\partial_{j}v_{2}-\partial_{2}v_{j}\partial_{j}v_{1}+v_{j}\partial_{j}\omega\\
 & = & \partial_{1}v_{1}\partial_{1}v_{2}+\partial_{1}v_{2}\partial_{2}v_{2}-\partial_{2}v_{1}\partial_{1}v_{1}-\partial_{2}v_{2}\partial_{2}v_{1}+v_{j}\partial_{j}\omega\\
 & = & \omega(\partial_{j}v_{j})+v_{j}\partial_{j}\omega\\
 & = & v_{j}\partial_{j}\omega\end{eqnarray}

Moreover, every incompressible vector field is of the form

\[
v_{1}=\partial_{2}\chi,\quad v_{2}=-\partial_{1}\chi\]

for some `stream function' $\chi.$ Vorticity is then

\[
\omega=-\partial_{1}^{2}\chi-\partial_{2}^{2}\chi\equiv\Delta\chi.\]

where $\Delta$ is a positive laplacian. Given appropriate boundary
conditions, this is an invertible operator. Thus we can regard the
vorticity as the dynamical variable and the velocity potential as
derived from it by solving the above elliptic differential equation.

Also,\[
v_{j}\partial_{j}\omega=\partial_{2}\chi\partial_{1}\omega-\partial_{1}\chi\partial_{2}\omega=\left\{ \chi,\omega\right\} \]

which is the Poisson bracket on a two dimensional phase space. It
is anti-symmetric and satisfies the Jacobi identity.Thus, two dimensional
incompressible flow reduces to the pair of equations

\[
\frac{\partial\omega}{\partial t}+\left\{ \chi,\omega\right\} =0,\quad\omega=\Delta\chi.\]

The Green's function of the laplacian (with appropriate boundary conditions),
can be thought of a linear operator which solves the Poisson equation
$\chi=K\omega$. Then the equation of motion of an incompressible
fluid in two dimensions becomes the non-linear integro-differential
equation

\begin{equation}
\frac{\partial\omega}{\partial t}+\left\{ K\omega,\omega\right\} =0.\label{eq:EulerVorticity-1}\end{equation}

\section{The Rigid Body}

It is useful to start with an example of a mechanical system that
looks like the opposite extreme from a fluid: a rigid body. We will
see that there are many similarities in the basic mathematical formulation,
although fluid mechanics is much more complicated. It is interesting
that the basic equations of both extremes are due to Euler.

\subsection{Euler Equations}

Recall the Euler equations of a rigid body on which no external forces
are acting are an expression of the conservation of angular momentum
$\mathbf{L}$. In the non-inertial reference attached to the body
itself, this takes the form

\[
\frac{d\mathbf{L}}{dt}+\mathbf{\Omega}\times\mathbf{L}=0\]

where $\mathbf{\Omega}$ is the angular velocity. These quantities
are related by a positive symmetric tensor (linear operator), $I$
the moment of inertia:

\[
\mathbf{L}=I\mathbf{\Omega}.\]

The moment of inertia $I$ is defined in terms of the density of the
rigid body as follows:

\[
I_{ij}=\delta_{ij}M_{kk}-M_{ij},\quad M_{ij}=\int\rho(x)x_{i}x_{j}dx,\]

It is obvious that the matrix $M\geq0$. Suppose its eigenvalues are
$M_{1},M_{2},M_{3},$ all positive numbers. Then $M_{kk}=M_{1}+M_{2}+M_{3}$
. The matrices $M$ and $I$ are diagonal in the same basis, and the
eigenvalues of $I_{ij}$ are \[
I_{1}=M_{2}+M_{3},\quad I_{2}=M_{1}+M_{3}\quad I_{3}=M_{1}+M_{2}.\]

Thus $I\geq0$ as well. The basis in which $I$ is diagonal forms
the principal axes of the body and $I_{1},I_{2},I_{3}$are called
the principal moments of inertia.

If we compare with the Euler equations in two dimensions, we see that
the vorticity is analogous to angular momentum and the velocity potential
analogous to angular velocity. Also, the elliptic differential operator
$\Delta$ above is analogous to the moment of inertia. The cross product
of vectors (which is anti-symmetric and satisfies the Jacobi identity)
corresponds to the Poisson bracket of functions.

\[
L\longleftrightarrow\omega,\quad\Omega\longleftrightarrow\chi,\quad I\longleftrightarrow\Delta,\times\longleftrightarrow\left\{ ,\right\} .\]
 Of course, the angular momentum has only three independent components
while vorticity belongs to an infinite dimensional space. So fluid
mechanics is much more complicated.

There is a co-ordinate system, which moves together with the body,
in which $I_{ij}$ is diagonal:

\[
L_{1}=I_{1}\Omega_{1},\quad L_{2}=I_{2}\Omega_{2},\quad L_{3}=I_{3}\Omega_{3}.\]

In terms of these the Euler equations become

\begin{equation}
\frac{d\Omega_{1}}{dt}+\frac{I_{2}-I_{3}}{I_{1}}\Omega_{2}\Omega_{3}=0,\quad\frac{d\Omega_{2}}{dt}+\frac{I_{3}-I_{1}}{I_{2}}\Omega_{3}\Omega_{1}=0\quad\frac{d\Omega_{3}}{dt}+\frac{I_{2}-I_{2}}{I_{3}}\Omega_{1}\Omega_{2}=0\label{eq:EulerJacobi}\end{equation}

They are solved in terms of the three Jacobi elliptic functions .

\subsubsection{Hamiltonian Formalism}

A rigid body moves so that the distance between any two points on
it remains fixed. Thus its configuration space is the isometry group
of $R^{3}.$ More precisely the connected component of the isometry
group, which consists of rotations and translations. The translations
are uninteresting as they simply describe a straight-line on which
the center of mass moves. Thus, we can think of the group of rotations
$SO(3)$ as the configuration space of a rigid body.

The components of angular momentum satisfy the Poisson bracket relations

\[
\left\{ L_{1},L_{2}\right\} =L_{2},\quad\left\{ L_{2},L_{3}\right\} =L_{3},\quad\left\{ L_{3},L_{1}\right\} =L_{1}.\]

Just as the kinetic energy due to translational motion of a particle
is $\frac{\mathbf{P}^{2}}{2m}$, the rotational kinetic energy of
a rigid body is given by

\begin{equation}
H=\frac{L_{1}^{2}}{2I_{1}}+\frac{L_{2}^{2}}{2I_{2}}+\frac{L_{3}^{2}}{2I_{3}}\label{eq:ellipsoid}\end{equation}

in the basis where $I_{ij}$ is diagonal. Indeed, we can check that
Euler equations above are implied by this hamiltonian and the above
Poisson brackets:

\[
\frac{dL_{i}}{dt}=\left\{ H,L_{i}\right\} .\]

\subsubsection{The Elliptic Curve}

The Poisson algebra of angular momentum has a non-trivial center.
That is, there is a polynomial in the generators that commutes with
(has zero Poisson brackets with) all the generators:

\begin{equation}
L^{2}=L_{1}^{2}+L_{2}^{2}+L_{3}^{2}.\label{eq:sphere}\end{equation}

It is the square of the magnitude of anguar momentum. In particular,
it commutes with the hamiltonian and hence is a conserved quantity:

\[
\frac{dL^{2}}{dt}=0.\]

Of course, the hamiltonian is itself a conserved quantity.

Thus the solution to the Euler equations describes parametrically
the curve which is the intersection of the sphere (\ref{eq:sphere})
with the ellipsoid (\ref{eq:ellipsoid}). This intersection can be
either a union of disjoint circles immersed in $R^{3}$ or a single
connected closed curve, depending on the values of $H$ and $L$.

So the solution of Euler's equation has a purely algebraic description.
When solving algebraic equations, it is useful to continue to complex
values, even if the physical values are real as in our case. The intersection
curve is then a complex curve, a manifold of two real dimension. A
moment's thought will convince you that this manifold must be a torus:
it is compact, and the time evolution defines an everywhere non-zero
vector field on it. (Only connectedness needs a proof which we skip.)
Thus was born the theory of elliptic curves.

The theory of elliptic curves have been honed into a fine marble sculpture
in the garden of mathematics. But some of the life has been lost in
this process: the physics seems to be lost.

\subsubsection{Geodesics on $SO(3)$}

Euler equations have a natural geometric interpretation as geodesics
on the rotation group, with respect to a metric determined by the
moment of inertia. This metric is, in interesting cases, not bi-invariant.
Instead, it is only invariant under the left action. Thus we can visualize
the universal cover of the rotation group as a three dimensional ellipsoid.

To see where this metric comes from, remember that the Lie algebra
of a group can be thought of as the space of left-invariant vector
fields. A positive quadratic form on the Lie algebra defines a left
invariant metric on the group. Euler's equations describe how the
tangent vectors to the geodesics evolve, with `time' having the meaning
of arc length.

This point of view will be very useful for us in understanding hydrodynamics.
So we will describe in more detail the geometry of left-invariant
metrics in a later section.

\section{Hamiltonian Systems From Lie Algebras}

We will see that the Euler equation of both ideal fluid mechanics
and the rigid body are special cases of a class of dynamical systems
obtained from a real Lie algebra $\mathcal{G}$. We digress a bit
to describe this class of systems.

The set of observables of a classical mechanical system is a \emph{Poisson
algebra.} That is, a commutative algebra on which is defined in addition
a bilinear (Poisson bracket) such that
\begin{enumerate}
\item $\left\{ f_{1},f_{2}\right\} =-\left\{ f_{2},f_{1}\right\} $ anti-symmetry
\item $\left\{ \left\{ f_{1},f_{2}\right\} ,f_{3}\right\} +\left\{ \left\{ f_{2},f_{3}\right\} ,f_{1}\right\} +\left\{ \left\{ f_{3},f_{1}\right\} ,f_{2}\right\} =0$
Jacobi identity
\item $\left\{ f_{1},f_{2}f_{3}\right\} =\left\{ f_{1},f_{2}\right\} f_{3}+f_{2}\left\{ f_{1},f_{3}\right\} $
Leibnitz identity
\end{enumerate}
The algebra of functions $f:\mathcal{G}^{*}\to R$ on the dual of
a Lie algebra is an example of a Poisson algebra. The exterior derivative
of such a function can be thought of as valued in the dual of $\mathcal{G}^{*}$,
which can be identified with $\mathcal{G}.$ Thus it makes sense to
take the Lie bracket of a pair of such exterior derivatives $[df_{1}(a),df_{2}(a)]$
evaluated at some point $a\in\mathcal{G}^{*}.$ A contraction with
$a$ itself gives a number. So we define the Poisson bracket to be

\[
\left\{ f_{1},f_{2}\right\} (a)=-i_{a}\left([df_{1}(a),df_{2}(a)]\right).\]
 The required properties follow from those of a Lie algebra and the
exterior derivative.

If we choose a particular function $H$ as the hamiltonian, we get
a dynamical system for which the time evolution of any observable
is given by

\[
\frac{df}{dt}=\left\{ H,f\right\} .\]

Now, every Lie algebra element $\omega\in\mathcal{G}$ defines a linear
function on its dual. In this case, we get a simpler form of this
equation: \[
\frac{d\omega}{dt}+[dH,\omega]=0.\]

\subsection{Metric Lie Algebra}

A Metric Lie Algebra is a Lie algebra along with an inner product
(usually not invariant). Now, an inner product is a symmetric tensor
on $\mathcal{G},$ and hence can be thought of as a quadratic function
on $\mathcal{G}^{*}.$ If we choose this function as the hamiltonian,
$dh:\mathcal{G}^{*}\to\mathcal{G}$ will be a linear function; which
is another way of thinking of an inner product.

If our Lie algebra $\mathcal{G}$ admits an invariant inner product
$<,>$ (not necessarily $H$), we can write this equation a bit more
explicitly. Then we can use the invariant inner product to identify
$\mathcal{G}^{*}$ and $\mathcal{G}$ and $dH$ becomes just a linear
operator$K:\mathcal{G}^{*}\to\mathcal{G}$ :

\[
H(\omega)=\frac{1}{2}<\omega,K\omega>.\]

The equations of motion are then

\[
\frac{d\omega}{dt}+[K\omega,\omega]=0.\]

An example of this is the rigid body, where the Lie algebra is the
cross product, the invariant inner product is the dot product and
$K$ is the inverse of the moment of inertia, yielding Euler's equations
.

When the Lie algebra admits no invariant inner product it is not as
easy to write an explicit form for the equations. It is then useful
to describe it in a basis $L_{i}$ of the Lie algebra. The Poisson
bracket is determined completely by its effect on the generators:

\[
\left\{ L_{i},L_{j}\right\} =c_{ij}^{k}L_{k}\]

where $c_{ij}^{k}$ are the structure constants of the Lie algebra
in this basis. Thus for example,

\[
\left\{ L_{i},L_{j}L_{m}\right\} =c_{ij}^{k}L_{k}L_{m}+c_{im}^{k}L_{j}L_{k}\]

We are thinking of the basis elements of $\mathcal{G}$ as generators
of the algebra of polynomials on $\mathcal{G}^{*}.$ This is a formal
translation of the point of view of most physicists.

A particularly simple polynomial is given by an inner product on the
Lie algebra. Expressed in the above basis, \[
H=\frac{1}{2}h^{ij}L_{i}L_{j}\]
 where $h^{ij}=h^{ji}.$ If we use this as a hamiltonian, the equations
of time evolution

\[
\frac{df}{dt}=\left\{ H,f\right\} .\]

can be written as

\[
\frac{dL_{k}}{dt}=h^{ij}\ c_{ik}^{m}\ L_{m}L_{j}.\]

If the inner product $h$ where invariant, it would satisfy the identity

\[
h^{ij}\ c_{ik}^{m}+i\leftrightarrow j=0\]

and there would be no time evolution.

\subsection{The Metric Lie Algebra of a Two Dimensional Incompressible Fluid}

The set of functions on the plane form a Lie algebra under the Poisson
bracket

\[
\left\{ f,g\right\} =\partial_{2}f\partial_{1}g-\partial_{1}f\partial_{2}g.\]

There an invariant inner product in this Lie algebra

\[
<f,g>=\int fgdx.\]

If choose as hamiltonian the positive quadratic form

\[
H(\omega)=\frac{1}{2}<\omega,G\omega>\]

where $G$ is the Green's function of the Laplace operator $\Delta$,
the equations of motion obtained are exactly those of an incompressible
fluid in two dimensions. %
\footnote{We must assume appropriate boundary conditions on $\omega$ so that
$H(\omega)$ exists.%
}

Thus the inner product that defines an incompressible fluid is

\[
\int f(x)G(x,y)g(y)dxdy\]

where

\[
\left(\partial_{1}^{2}+\partial_{2}^{2}\right)G(x,y)=-\delta(x,y)\]

with appropriate boundary conditions.

\subsection{Dimensions Greater Than Two}

Even if the dimension is greater than two, the hamiltonian of incompressible
fluid is still a quadratic function on the Metric Lie Algebra of incompressible
vector fields. However there is no longer any obvious invariant inner
product, so we have to contend with describing the equations a bit
indirectly.

The condition of incompressibility

\begin{equation}
\partial_{k}\left(\rho v^{k}\right)=0\label{eq:incompressible}\end{equation}

is equivalent to \[
\mathcal{L}_{v}\rho=0.\]

It follows that the commutator of two incompressible vector fields
is also incompressible: the set of solutions $\mathcal{V}_{\rho}$
of \ref{eq:incompressible} is a Lie algebra with the same Lie bracket
as before. A linear function on $\mathcal{V}_{\rho}$ is of the form
$\int\rho v^{i}a_{i}dx$ for some 1-form $a$. But, a gauge transformation
\[
a\mapsto a+d\Lambda\]

leaves this function unchanged. Thus the dual of $\mathcal{V}_{\rho}$
is the space of 1-forms modulo exact 1-forms:

\[
\mathcal{V}_{\rho}^{*}\equiv\Lambda^{1}/d\Lambda^{0}.\]

The kinetic energy of the fluid is a simple quadratic form on the
Lie algebra of incompressible vecor fields.: $\frac{1}{2}\int\rho v^{i}v^{j}g_{ij}dx.$
The dual variable dual to $v^{i}$ can be thought of as just

\[
a_{i}=[g_{ij}v^{j}]\]

the square brackets being there to remind us that we must take the
equivalence classes under gauge transformations. Thus the Hamiltonian
can only depend on the vorticity

\[
\omega=da,\quad\omega_{ij}=\partial_{i}a_{j}-\partial_{j}a_{i}.\]

But it must be a zeroth order operator in terms of velocity.

\[
H=\frac{1}{2}(\omega,G\omega)\]

where $G$ is the Green's function of the elliptic system

\[
\omega=da,\quad\partial_{i}\left(\rho g^{ij}a_{j}\right)=0.\]

The Euler equations in vorticity form are

$\frac{\partial\omega}{\partial t}+\mathcal{L}_{v}\omega$=0

where we are to regard $v$ as the unique incompressible vector field
determined by the vortcity by

\[
\partial_{i}[\rho v^{i}]=0,\quad\partial_{i}[g_{jk}v^{k}]-\partial_{j}[g_{ik}v^{k}]=\omega_{ij}\]

with appropriate boundary conditions.

\section{Three Dimensional Incompressible Flow}

This is the most important kind fluid flow: the vast majority of physical
phenomena take place in three dimensions and the velocities are small
compared to the speed of sound.

\subsection{The Clebsch Variables}

In this case Euler equations can be expressed in a simpler form using
a parametrization due to Clebsch \cite{Lamb}:

\begin{equation}
\omega=dq\wedge dp\label{eq:Clebsch}\end{equation}

It is clear that this is only possible because $d\omega=0.$Locally,
any one-form in $R^{3}$ can be expressed in the form \[
a=d\lambda+qdp\]

so that any exact two form is locally of the form \ref{eq:Clebsch}.

In terms of these variables, Euler equations become the statement
that $p,q$ are constant along streamlines:

\[
\frac{\partial p}{\partial t}+v^{i}\partial_{i}p=0,\quad\frac{\partial q}{\partial t}+v^{i}\partial_{i}q=0\]
 The velocity is determined in terms of $p,q$ by

\[
v^{i}=g^{ij}[\partial_{j}\lambda+q\partial_{j}p]\]

Here, $\lambda$ is eliminated by the condition of incompressibility

\[
\partial_{i}[\rho v^{i}]=0\iff\partial_{i}[\rho g^{ij}(\partial_{j}\lambda+q\partial_{j}p)]=0.\]

\subsubsection{Canonical Relations}

The Clebsch variables $p,q$ are canonical conjugates of each other.
That is, if we postulate canonical commutation relations\[
\left\{ p(x),q(y)\right\} =\delta(x-y),\quad\left\{ p(x),p(y)\right\} =0=\left\{ q(x),q(y)\right\} \]
 and the hamiltonian

\[
H=\frac{1}{2}\int\rho v^{i}v^{j}g_{ij}dx\]

with $v^{i}$determined in terms of $p,q$ as above, we get the Clebsch
form of the Euler equations.

To see this note that the canonical commutation relations give

\[
\left\{ F,p(y)\right\} =-\frac{\delta F}{\delta q(y)},\quad\left\{ F,q(y)\right\} =\frac{\delta F}{\delta p(y)}\]

for any function of $p,q$. For example, if

\[
F=\int f^{ij}\partial_{i}q\partial_{j}pdx\]

\begin{equation}
\left\{ F,p(y)\right\} =-\partial_{j}f^{ij}\partial_{i}p,\quad\left\{ F,q(y)\right\} =-\partial_{j}f^{ij}\partial_{i}q\label{eq:VariationF}\end{equation}

if $f^{ij}$is independent of $p,q.$ Now, the kinetic energy of the
fluid can also be writen as

\[
H=\frac{1}{2}\int\rho v^{i}v^{j}g_{ij}dx=\frac{1}{4}\int f^{ij}\omega_{ij}dx=\frac{1}{2}\int f^{ij}\partial_{i}q\partial_{j}p\]

where $\rho v^{i}=\partial_{j}f^{ij}$and $f^{ij}=-f^{ji}$ is the
velocity potential. Therefore

\[
\frac{\partial p}{\partial t}=\left\{ H,p\right\} =-v^{i}\partial_{i}p,\quad\frac{\partial q}{\partial t}=\left\{ H,q\right\} =-v^{i}\partial_{i}q\]

which are the Euler equations.

An extra factor of 2 appears (cancelling the $\frac{1}{2}$ in the
Hamiltonian) compared to \ref{eq:VariationF} because $f^{ij}$itself
depends linearly on $v$ and hence on $p,q$.

\subsubsection{The Moment Map}

Marsden and Weinsetin \cite{MarsdenWeinstein} showed that the Clebsch
parametrization has a natural geometric interpretaion. On the space
$\mathcal{F}$ of real valued functions on a manifold with density,
there is an inner product\[
<f,g>=\int fg\rho dx.\]

This can be used to turn $\mathcal{F\oplus\mathcal{F}}$ into a symplectic
vector space: the conjugate pairs of functions $p,q$ parametrize
this phase space. Since volume preserving diffeomorphisms preserve
the inner product above, they must act as canonical transformation
on this phase space. The infinitesimal generaror these transformations
is just vorticity. The Clebsch parametrization

\[
\omega=dq\wedge dp\]

is analogous to the formula for angular momentum

\[
\mathbf{L}=\mathbf{r}\times\mathbf{p}\]

in Euclidean space. The dot product is a rotation invariant inner
product in $R^{3}$, which turns $R^{3}\oplus R^{3}$ into a phase
space on which the rotations act as canonical transformations generated
by $\mathbf{L}.$

\subsubsection*{A direction for further research}

This point of view allows us to generalize the theory of a three dimensional
incompressible fluid to the case where the underlying manifold is
non-commutative sphere: a kind of `regularization' of hydrodynamics
where each point is replaced by a fuzzy object which is the average
of many points. In addition to being more mathematically rigorous
(the equations of motion are finite dimensional ODEs instead of PDEs)
this might be a physically realistic description of the large scale
behavuor of fluids: the small scale fluctuations in velocity are averaged
out to get an effective theory that s not no longer local.

\subsection{The Lie Algebra of an Ideal Isentropic Fluid}

The set of vector fields $\mathcal{V}$ on a manifold is a Lie algebra
under the commutator or Lie bracket,\[
[u,v]^{i}=u^{k}\partial_{k}v^{i}-v^{k}\partial_{k}u^{i}.\]

$\mathcal{V}$ acts on the space of scalars $\mathcal{F}$ through
the derivative

\[
\mathcal{L}_{u}\phi=v^{k}\partial_{k}\phi.\]

The sum of the two vectors spaces $\mathcal{V}\oplus\mathcal{F}=\mathcal{G}$
is thus a Lie algebra as well:

\[
\left[(u,\phi),(\tilde{u},\tilde{\phi})\right]=\left([u,\tilde{u}],\mathcal{L}_{u}\tilde{\phi}-\mathcal{L}_{\tilde{u}}\phi\right)\]

This is the semi-direct sum of the Lie algebra of vector fields and
the abelian Lie algebra of scalars.

The dual of $\mathcal{F}$ is the space of scalar densities $\mathcal{F}^{*}$:

\[
i_{\phi}\rho=\int\phi\rho dx.\]

The dual of $\mathcal{V}$ is the space of co-vector densities $\mathcal{V}^{*}$:

\[
i_{v}j=\int v^{i}j_{i}dx.\]

Given functions $F,G:\mathcal{V}^{*}\oplus\mathcal{F}^{*}\to R$,
their Poisson brackets are given by

\[
\left\{ F,G\right\} =\int\rho\left(\frac{\delta F}{\delta j_{k}}\partial_{k}\frac{\delta G}{\delta\rho}-\frac{\delta G}{\delta j_{k}}\partial_{k}\frac{\delta F}{\delta\rho}\right)dx+\int j_{i}\left(\frac{\delta F}{\delta j_{k}}\partial_{k}\frac{\delta G}{\delta j_{i}}-\frac{\delta G}{\delta j_{k}}\partial_{k}\frac{\delta F}{\delta j_{i}}\right)dx\]

This can be written also as

\[
\left\{ F,G\right\} =\int\frac{\delta G}{\delta j_{k}}\left[-\rho\partial_{k}\frac{\delta F}{\delta\rho}-j_{i}\partial_{k}\frac{\delta F}{\delta j_{i}}-\partial_{i}\left(j_{k}\frac{\delta F}{\delta j_{i}}\right)\right]dx-\int\frac{\delta G}{\delta\rho}\partial_{k}\left[\rho\frac{\delta F}{\delta j_{k}}\right]dx.\]

Choosing $G$ to be a linear function, we get a particularly convenient
form of the Poisson bracket:

\[
\left\{ F,\ j_{k}\right\} =-\rho\partial_{k}\frac{\delta F}{\delta\rho}-j_{i}\partial_{k}\frac{\delta F}{\delta j_{i}}-\partial_{i}\left(j_{k}\frac{\delta F}{\delta j_{i}}\right)\]

\[
\left\{ F,\ \rho\right\} =-\partial_{k}\left[\rho\frac{\delta F}{\delta j_{k}}\right]\]

So far we have not needed any additional geometric structures such
as a Riemannian metric: all the derivatives above make sense without
the need of a connection.

Now we identify the variables $j_{i},\rho$ physically as momentum
density and mass density respectively. The Hamiltonian is the sum
of kinetic and potential (internal) energies of the fluid:

\[
H=\int\frac{j_{i}j_{j}}{2\rho}g^{ij}dx+\int U(\rho)dx\]

where $g_{ij}$ is a Riemannian metric on the underlying space.

On physical grounds we know also that momentum density is mass density
times velocity:

\[
j_{i}=g_{ik}\rho v^{k}\]

so that

\[
\frac{\delta H}{\delta j_{i}}=v^{i}.\]

The equations of motion

\[
\frac{\partial j_{i}}{\partial t}=\left\{ H,j_{i}\right\} ,\quad\frac{\partial\rho}{\partial t}=\left\{ H,\rho\right\} \]

become

\[
\frac{\partial j_{i}}{\partial t}=-\rho\partial_{k}\left[-\frac{1}{2}\frac{g^{ij}j_{i}j_{j}}{\rho^{2}}+\frac{\partial U}{\partial\rho}\right]-j_{i}\partial_{k}v^{i}-\partial_{i}\left(j_{k}v^{i}\right)\]

\[
\frac{\partial\rho}{\partial t}=-\partial_{k}\left(\rho v^{k}\right).\]

We can simplify the first of these

\[
\frac{\partial j_{i}}{\partial t}+\nabla_{i}\left(j_{k}v^{i}\right)=\frac{1}{2}\rho\partial_{k}\left[g_{ij}v^{i}v^{j}\right]-\rho v^{i}g_{ij}\nabla_{k}v^{j}-\rho\partial_{k}\left[\frac{\partial U}{\partial\rho}\right]\]

where $\nabla_{i}$ is the covariant derivative of the Riemann metric
$g_{ij}.$The first two terms on the right hand side cancel each other.
If we identify the enthalpy as

\[
w=\frac{\partial U}{\partial\rho}\]

we get

\[
\frac{\partial j_{i}}{\partial t}+\nabla_{i}\left(j_{k}v^{i}\right)+\rho\partial_{k}w=0\]

or

\[
\frac{\partial v^{i}}{\partial t}+v^{k}\nabla_{k}v^{i}+g^{ik}\partial_{k}w=0\]

along with the conservation of mass:

\[
\frac{\partial\rho}{\partial t}+\partial_{k}\left(\rho v^{k}\right)=0.\]

These are exactly the Euler equations for an isentropic ideal fluid
we obtained earlier, except that there we looked at the special case
of the Euclidean metric. The Lie algebra $\mathcal{V}\oplus\mathcal{F}$
as well as the Poisson algebra following from it are independent of
the choice of metric. But the hamiltonian $H$ depends on this choice.
In this case, the hamiltonian is not a quadratic function, unlike
in the case of the rigid body or the incompressible fluid. If the
equation of state is a polytrope $w=\rho^{3}$ it is possible to think
of the hamiltonian as a cubic function. Perhaps this is worth deeper
study.

\section{Riemannian Geometry}

We review here some facts about Riemannian manifolds, to rephrase
some basic facts in a language convenient for our purposes. This is
not meant as an introduction to Riemannian geometry. There are several
excellent texts available, in particular the one by Chavel \cite{Chavel}.

\subsection{Covariant Derivative}

A covariant derivative (connection) $\nabla_{u}v$ of a vector field
$v$ on a manifold $M$ along another vector field $u$ should satisfy
the conditions of linearity in $u,v$ as well as

\[
\nabla_{u}[fv]=f\nabla_{u}v+u(f)v,\quad\nabla_{fu}v=f\nabla_{u}v.\]

Thus it involves first derivatives of $v$ and no derivative of $u.$
Explictly in co-ordinates

\[
[\nabla_{u}v]^{i}=u^{j}\partial_{j}v^{i}+\Gamma_{jk}^{i}u^{j}v^{k}\]

for a set of connection coefficients $\Gamma_{jk}^{i}$. We will only
be interested in connections without torsion:

\[
\nabla_{u}v-\nabla_{v}u=[u,v].\]

The cuvature is the tensor field defined by

\[
R(u,v)w=\nabla_{[u,v]}w-\nabla_{v}\nabla_{u}w+\nabla_{u}\nabla_{v}w.\]

Given a Riemannian metric $g$ on the manifold, there is a unique
connection of zero torsion and which preserves the metric:

\[
g(\nabla_{u}v,w)+g(v,\nabla_{u}w)=u\left(g(v,w)\right).\]

Explicitly in co-ordinates, this connection has coefficients

\[
\Gamma_{jk}^{i}=\frac{1}{2}g^{il}\left\{ \partial_{j}g_{lk}+\partial_{k}g_{lj}-\partial_{l}g_{jk}\right\} \]

The curves that extermize the action

\[
\int g(\dot{x},\dot{x})dt\]

are the geodesics; they satisfy the equation

\[
\nabla_{\dot{x}}\dot{x}=0\]
 or

\[
\ddot{x}^{i}+\Gamma_{jk}^{i}\dot{x^{j}}\dot{x}^{k}=0.\]

It will be convenient to define the curvature bi-quadratic form

\[
R(u,v)=g\left(R(u,v)v,u\right)\]

\subsection{Geodesic Deviation and Curvature}

The curvature form determines the rate of deviation of nearby geodesics
from each other.

Consider a geodesic $x^{i}(t)$ , expressed in a co-ordinate system
centered at the initial point: $x^{i}(0)=0.$

\[
\frac{d^{2}x^{i}}{dt^{2}}+\Gamma_{jk}^{i}\frac{dx^{j}}{dt}\frac{dx^{k}}{dt}=0.\]

\[
\frac{D\dot{x}}{dt}=0.\]

Let the initial velocity be $v\in T_{0}M.$ An infinitesimally close
geodesic to this one will satisfy

\[
\frac{D^{2}y^{i}}{dt^{2}}+R_{jkl}^{i}\frac{dx^{j}}{dt}\frac{dx^{k}}{dt}y^{l}=0.\]

\[
\frac{D^{2}y}{dt^{2}}+R(\dot{x},y)\dot{x}=0.\]

The vector field $y$ along the original geodesic connects the points
at equal time on two nearby geodesics. It is called the Jacobi vector
field. We can derive an equation for the length squared of the Jacobi
field \[
\frac{1}{2}\frac{d^{2}|y(t)|^{2}}{dt^{2}}=\left|\frac{Dy}{dt}\right|^{2}-R(\dot{x},y).\]

Suppose the initial conditions for the Jacobi equation are

\[
y(0)=0,\quad\frac{Dy^{i}}{dt}(0)=u,\quad\dot{x}=v.\]

That is, we consider two geodesics starting at the same point but
with slightly different initial velocities.

Then

\[
|y(t)|^{2}=t^{2}|u|^{2}-\frac{t^{3}}{3}R(u,v)+\mathrm{O}(t^{4}).\]

The first term would have been the answer in Euclidean space. Thus,
if $R(u,v)<0$, the geodesic with tangent vector $v$ is unstable
with respect to an infinitesimal perturbation in the direction $u$.

\subsection{Curvature as a Biquadratic}

We saw that the curvature tensor on a Riemannian manifold describes
the behavior of geodesics under small changes of the initial conditions.
Therefore, it controls the stability properties of a physical system
whose time evolution is along geodesics. It will be useful to calculate
this tensor for left-invariant metrics on a Lie group to understand
the stability of systems such as the rigid body or an ideal fluid.
The formulas can get quite complicated, but a trick mentioned in Milnor's
article allows us a simpler description. We will derive a simple formula
that is quite useful in our applications by going a little beyond
Milnor in this direction.

Recall that a co-variant symmetric tensor on a vector space is exactly
the same thing as a quadratic form. A covariant symmetric tensor is
a bilinear map $Q:V\times V\to R$, satisfying $Q(u,v)=S(v,u).$ A
quadratic form is a function $Q:V\to R$ that satisfies the scaling
property

\[
Q(\lambda u)=\lambda^{2}Q(u).\]

Given a covariant symmetric tensor we can construct a quadratic by
taking the special case when its entries are equal:

\[
Q(u)=Q(u,u).\]

Conversely, we can get a symmetric tensor from a quadratic by `polarization':

\[
Q(u,v)=\frac{Q(u+v)-Q(u)-Q(v)}{2}.\]

These maps are inverses of each other.

In the same spirit, a tensor with the symmetries of the Riemmanian
curvature is fully determined by a bi-quadratic form on the tangent
space. In the applications we have in mind, this is a much more convenient
description, as we will be able to write expicit formulas and identify
their positivity properties more easily.

Recall that a bi-quadratic is a function on a vector space $T:V\times V\to R$
that satisfies

\[
T(\lambda u,v)=\lambda^{2}T(u,v),\quad T(u,\lambda v)=\lambda^{2}T(u,v).\]

Now, the Riemann curvature tensor is a fourth rank tensor defined
by

\[
r(u,v,w,x)=g\left({\nabla}_{[u,v]}w-{\nabla}_{u}{\nabla}_{v}w+{\nabla}_{v}{\nabla}_{u}w,x\right).\]

It defines a biquadratic on the tangent space by choosing $w=u,v=x$
\cite{Milnor}

\[
r(u,v)=g\left({\nabla}_{[u,v]}u,v\right)-G\left({\nabla}_{u}{\nabla}_{v}u,v\right)+G\left({\nabla}_{v}{\nabla}_{u}u,v\right)\]

This function satisfies \begin{equation}
r(\lambda u,v)=\lambda^{2}r(u,v),\quad r(u,v)=r(v,u),\quad r(u.u)=0.\label{eq:biquardaticidentities}\end{equation}

The meaning of these conditions is that $k(u,v)=\frac{r(u,v)}{g(u,u)g(v,v)-g(u,v)^{2}}$
depends on the subspace defined by $u,v$. To prove this, we just
have to show that the condition

\begin{equation}
r(au+bv,cu+dv)=(ad-bc)^{2}r(u,v)\label{eq:biquadraticidentity}\end{equation}

is equivalent to the conditions (\ref{eq:biquardaticidentities}).
The denominator is the area of the parallelogram in the tangent space
with sides $u,v.$ The ratio $k(u,v)$ is the sectional curvature
of the plane spanned by $u,v.$

As Milnor points out, the conditions (\ref{eq:biquardaticidentities})
imply in turn all the symmetry properties of a Riemann tensor. Also,
$k(u,v)$ determines fully the Riemann curvature tensor through a
`polarization' identity\cite{Chavel}:

\[
r(u,v,w,x)=\frac{1}{6}\left[\frac{\partial^{2}}{\partial s\partial t}\left\{ r(u+sw,v+tx)-r(u+sx,v+tw)\right\} \right]_{s=t=0}.\]

The Ricci tensor is the trace of the Riemann tensor; equivalent to
it is the quadratic form

\[
r(u)=r(u,e_{i},u,e_{j})g^{ij}\]

where $e_{i}$is some basis in the tangent space and $g^{ij}$is the
inverse of the matrix of inner products $g_{ij}=g(e_{i},e_{j}).$

From our current point of view it can be viewed as the average over
all vectors

\[
r(u)=\int r(u,v)d\mu_{g}(v).\]

The average is with respect to the Gaussian measure on the tangent
space, with zero mean and covariance given by the metric g. Recall
that in terms of components

\[
\int v^{i}v^{j}d\mu_{g}(v)=g^{ij}\]

Also, the Ricci scalar is the further average over $u$:

\[
r=\int r(u)d\mu_{g}(u).\]

This point of view has the advantage that it extends to infinite dimensions.
Also it is more natural in applications involving random forces.

\section{Geometry of Left-Invariant Metrics }

Let $\mathfrak{G}$ be a Lie group. Its Lie algebra $\mathcal{G}$
can be thought of either as the tangent space at the identity or as
the space of left invariant vector fields on $\mathfrak{G}.$ An inner
product $G$ on $\mathcal{G}$ is thus equivalent to a left-invariant
Riemannian metric on $\mathfrak{G}.$ We will study the geodesics
and curvature of this metric. Using homogenity, all computations can
be reduced to the Lie algebra. The basic reference is the article
by Milnor \cite{Milnor}, especially Section 5. We will go beyond
Milnor in deriving explicit formulas in a form useful for applications
to mechanics.

\subsection{Covariant Derivative}

The covariant derivative (Levi-Civita connection) is determined by
the derivative of a left-invariant vector field by another. We will
denote the covariant derivative on a group manifold by $D$ to distinguish
it from the covariant derivative on a general Riemannian manifold,
which we denote by $\nabla$. This will be helpful when we talk of
the diffeomorphism group of a Riemannin manifold: the derivative on
the group of diffeomorphisms and that on the underlying manifold are
closely related, but not identical notions.

The conditions of zero torsion and preserving the metric become

\[
{D}_{u}v-{D}_{v}u=[u,v],\]
 \[
G\left({D}_{u}v,w\right)+G\left(v,{D}_{u}w\right)=0;\]

The zero torsion condition can also be written as

\[
G\left({D}_{u}v,w\right)-G\left({D}_{v}u,w\right)=G\left([u,v],w\right).\]

Its cyclic permutations give

\[
G\left({D}_{v}w,u\right)-G\left({D}_{w}v,u\right)=G\left([v,w],u\right)\]

\[
G\left({D}_{w}u,v\right)-G\left({D}_{u}w,v\right)=G\left([w,u],v\right)\]

Adding the first and third and subtracting the second, and using the
invariance of the metric

\[
G\left({D}_{u}v,w\right)=\frac{1}{2}\left\{ G\left([u,v],w\right)-G\left([v,w],u\right)+G\left([w,u],v\right)\right\} \]

Define the linear operator $\tilde{u}:\mathcal{G}\to\mathcal{G}$
by

\begin{equation}
G\left(\tilde{u}v,w\right)=G\left([u,v],w\right)+G\left(v,[u,w]\right).\label{eq:deformation}\end{equation}

If the metric $G$ were invariant under the Lie algebra, $\tilde{u}$
would vanish for all $u.$ Thus it measures the deformation of the
metric. Now,

\begin{equation}
G\left(\tilde{u}v,w\right)+G\left(\tilde{v}u,w\right)=G\left([u,w],v\right)+G\left([v,w],u\right).\label{eq:deformation2}\end{equation}

Thus

\[
{D}_{u}v=\frac{1}{2}\left\{ [u,v]-\tilde{u}v-\tilde{v}u\right\} .\]

\subsection{Geodesics on a Group Manifold}

Given a curve $\gamma:[a,b]\to\mathfrak{G}$, its tangent vector at
each point can be thought of as an element of the Lie algebra:

\[
\frac{d\gamma}{dt}=\gamma v.\]

Thus the tangent vectors give a curve in the Lie algebra $v:[a,b]\to\mathcal{G}.$

We can defines the action of the curve as

\[
S(\gamma)=\frac{1}{2}\int_{a}^{b}G(v,v)dt.\]

(Some geometers call this quantity the energy; physicists would call
it the action.)

A geodesic is a curve at which the action is stationary with respect
to small variations.

To compute this variation, let us consider a one parameter family
of curves; that is a map

$\phi:[0,\epsilon]\times[a,b]\to\mathfrak{G},$ for some positive
number $\epsilon.$ We require that the initial and final points are
left unchanged $\phi(s,a)$ and $\phi(s,b)$ are independent of $s.$

Define

\[
\frac{\partial\phi}{\partial t}=\phi v,\quad\frac{\partial\phi}{\partial s}=\phi u.\]

So $u(s,a)=u(s,b)=0.$

We get the integrability condition

\[
\partial_{s}v=\partial_{t}u+[v,u].\]

Regarding $\phi(.,s)$for each value of $s$ as a curve, the action
becomes a function of $s.$ Its derivative is

\[
\partial_{s}S(\gamma_{s})=\int_{a}^{b}G(\partial_{s}v,v)dt=\int_{a}^{b}G\left(\partial_{t}u+[v,u],v\right)dt\]

Using the linear operator $\tilde{v}:\mathcal{G}\to\mathcal{G}$ defined
earlier

\[
G\left([v,u],v\right)=G\left(u,\tilde{v}v\right).\]

By integration by parts

\[
\partial_{s}S(\gamma_{s})=-\int_{a}^{b}G(u,\partial_{t}v-\tilde{v}v)dt.\]

Thus the geodesic equation on the group becomes the ODE on the Lie
algebra

\begin{equation}
\partial_{t}v+{D}_{v}v=0.\label{eq:GeodesicEqnAlgebraic}\end{equation}

\subsection{Curvature of a Left-Invariant Metric}

We will now calculate explictly the curvature bi-quadratic for a Metric
Lie Algebra; i.e., for a Lie algebra $(\mathcal{G},G)$ with an inner
product on it.

\[
G\left({D}_{v}{D}_{u}u,v\right)=-G({D}_{u}u,{D}_{v}v)=-G\left(\tilde{u}u,\tilde{v}v\right).\]

\[
-G\left({D}_{u}{D}_{v}u,v\right)=G\left({D}_{v}u,{D}_{u}v\right)=-\frac{1}{4}|[u,v]|^{2}+\frac{1}{4}|\tilde{u}v+\tilde{v}u|^{2}.\]

\[
G\left({D}_{[u,v]}u,v\right)=\frac{1}{2}G\left([[u,v],u],v\right)-\frac{1}{2}|[u,v]|^{2}+\frac{1}{2}G\left([v,[u,v]],u\right)\]

Thus

\[
R(u,v)=-\frac{3}{4}|[u,v]|^{2}+\frac{1}{2}\left\{ G\left([[u,v],u],v\right)+G\left([v,[u,v]],u\right)\right\} +\frac{1}{4}|\tilde{u}v+\tilde{v}u|^{2}-G\left(\tilde{u}u,\tilde{v}v\right)\]

The middle term can be further simplified using

\[
G([u,w],v)=G(\tilde{u}w,v)-G(w,[u,v])\]

and $G(\tilde{u}w,v)=G(\tilde{u}v,w).$ We get

\[
\frac{1}{2}\left\{ G\left([[u,v],u],v\right)+G\left([v,[u,v]],u\right)\right\} =|[u,v]|^{2}+\frac{1}{2}G\left([u,v],\tilde{v}u-\tilde{u}v\right).\]

Thus

\[
R(u,v)=\frac{1}{4}|[u,v]|^{2}+\frac{1}{2}G\left([u,v],\tilde{v}u-\tilde{u}v\right)+\frac{1}{4}|\tilde{u}v+\tilde{v}u|^{2}-G\left(\tilde{u}u,\tilde{v}v\right)\]

If the metric is bi-invariant, only the first term survives: the curvature
of a bi-invariant metric is postive. We can `complete the square'
on the first two terms to get

\begin{equation}
R(u,v)=\frac{1}{4}|[u,v]+\tilde{v}u-\tilde{u}v|^{2}+G(\tilde{u}v,\tilde{v}u)-G\left(\tilde{u}u,\tilde{v}v\right)\label{eq:CurvatureG}\end{equation}

This simple formula appears to be a new result.

We already see something important: the sectional curvature in any
plane that contains a Killing vector is positive:

\[
\tilde{u}=0\implies R(u,v)=\frac{1}{4}|[u,v]+\tilde{v}u|^{2}.\]

This result is known but is hard to see with the formulas known in
the literature.

\subsection{Example: The Two Dimensional Lie Algebra}

The only non-abelian Lie algebra of dimension two has commutation
relations in terms of basis vectors

\[
[e_{0},e_{1}]=e_{1}\]

or in terms of components \[
[u,v]=\left(0,u_{0}v_{1}-v_{0}u_{1}\right).\]

The only metric is (up to a change of basis that does not change the
above commutation relations)

\[
G(u,v)=u_{0}v_{0}+u_{1}v_{1}.\]
 The corresponding group manifold is the half plane,

\[
H=\left\{ (x_{0},x_{1})|x_{0}>0\}\right\} \]

with the multiplication law

\[
(x_{0},x_{1})(x_{0}',x_{1}')=(x_{0}x_{0}',x_{0}x_{1}'+x_{1}).\]

The corresponding left-invariant metric is the Poincare metric. It
is well known that this metric has constant negative curvature. Hence
it can serve as a simple model of an unstable dynamical system.

We get

\[
G(\tilde{v}v,w)=G(v,[v,w])=v_{1}(v_{0}w_{1}-w_{0}v_{1})\]

so that

\[
\tilde{v}v=(-v_{1}^{2},v_{0}v_{1}).\]

The geodesic equation in the Lie algebra becomes

\begin{equation}
\frac{dv_{0}}{dt}=-v_{1}^{2},\quad\frac{dv_{1}}{dt}=v_{0}v_{1}\label{eq:geodesichalfplane1}\end{equation}

The solution is (scaling $t$ so that $|v|=1$)

\[
v_{0}=-\tanh t,\quad v_{1}=\frac{1}{\cosh t}.\]

To get the curve in the group we must solve

\[
\frac{dg}{dt}=gv\]

where $gv$ denotes the action of the group on its Lie algebra.

Viewing $v$ as an element infinitesimally close to the identity

\[
(x_{0},x_{1})v=(x_{0}v_{0},x_{0}v_{1})\]

Thus

\begin{equation}
\frac{dx_{0}}{dt}=x_{0}v_{0},\frac{dx_{1}}{dt}=x_{0}v_{1}.\label{eq:geodesichalfplane2}\end{equation}

The Riemannian metric is thus the Poincare' metric

\[
dt^{2}=\frac{dx_{0}^{2}+dx_{1}^{2}}{x_{0}^{2}}.\]

Thus geodesics are

\[
(\frac{A}{\cosh(t-t_{0})},B+A\tanh[t-t_{0}]).\]

(It is convenient to choose the origin of time as the point where
$v_{0}(t)$ vanishes instead of the starting time.) These are the
well-known semicircles

\[
x_{0}^{2}(t)+(x_{1}(t)-B)^{2}=A^{2}.\]

\subsubsection{Conserved Quantities of Geodesic Motion}

The Killing vectors of the Poincare' metric

\[
f_{0}=x_{0}\partial_{0}+x_{1}\partial_{1},\quad f_{1}=\partial_{1},\quad f_{-1}=(x_{1}^{2}-x_{0}^{2})\partial_{1}+2x_{0}x_{1}\partial_{0}\]

form the $\mathcal{SL}(2,R)$ Lie algebra:

\[
[f_{0},f_{1}]=f_{1},\quad[f_{0},f_{-1}]=f_{-1},\quad[f_{1},f_{-1}]=2f_{0}.\]

The corresponding conserved quantities of geodesic motion are

\[
F_{0}=x_{0}p_{0}+x_{1}p_{1},\quad F_{1}=p_{1},\quad F_{-1}=(x_{1}^{2}-x_{0}^{2})p_{1}+2x_{0}x_{1}p_{0}\]

where $p_{a}$ are the canonical conjugates of $x^{a}:$

\[
\left\{ p_{a},x_{b}\right\} =\delta_{ab}.\]

Since the geodesic equations have as hamiltonian

\[
H=\frac{1}{2}g^{ab}(x)p_{a}p_{b}\]
 we get $\frac{dx_{a}}{dt}=g^{ab}p_{b}.$ Expressing in terms of the
variables $v_{a}$introduced earlier

\[
p_{a}=\frac{v_{a}}{x_{0}}.\]
 Thus we see that the conserved quantities of the equations (\ref{eq:geodesichalfplane1},\ref{eq:geodesichalfplane2})
above are

\[
F_{0}=v_{0}+\frac{x_{1}}{x_{0}}v_{1},\quad F_{1}=\frac{v_{1}}{x_{0}},\quad F_{-1}=\frac{x_{1}^{2}-x_{0}^{2}}{x_{0}}v_{1}+2x_{1}v_{0}.\]

Of these, $F_{0},F_{1}$ generate the left translations of the half-plane
thought of as a group. Now, note that $F_{0}^{2}-F_{1}F_{-1}=v_{0}^{2}+v_{1}^{2}$
which is an `obvious' conserved quantity. Putting in the formula for
the $L$'s we get the semi-circle for the shape of the geodesic.

\subsubsection*{Curvature Form}

It is straightforward to calculate the curvature biquadratic for this
case:

\[
R(u,v)=-[u_{0}v_{1}-v_{0}u_{1}]^{2}\]

which simply says that the sectional curvature is:

\[
K(u,v)=-1.\]

We use this simple case to check the numerical factors in our formula
for curvature.

\subsection{Geodesics on $SO(3)$}

Consider $R^{3}$ as a Lie algebra with the cross-product as the Lie
bracket:

\[
[u,v]=\left(u_{2}v_{3}-u_{3}v_{2},u_{3}v_{1}-u_{1}v_{3},v_{2}-u_{2}v_{1}\right).\]
 A corresponding Lie group is $SO(3).$ Any inner product in $R^{3}$
can be brought to the diagonal form by a rotation without changing
the Lie bracket:

\[
G(u,v)=G_{1}u_{1}v_{1}+G_{2}u_{2}v_{2}+G_{3}u_{3}v_{3}\]

Thus

\[
G(\tilde{u}v,w)=G_{1}\left\{ \left[u_{2}v_{3}-u_{3}v_{2}\right]w_{1}+\left[u_{2}w_{3}-u_{3}w_{2}\right]v_{1}\right\} \]

\[
+G_{2}\left\{ \left[u_{3}v_{1}-u_{1}v_{3}\right]w_{2}+\left[u_{3}w_{1}-u_{1}w_{3}\right]v_{2}\right\} \]

\[
+G_{3}\left\{ \left[u_{1}v_{2}-u_{2}v_{1}\right]w_{3}+\left[u_{1}w_{2}-u_{2}w_{1}\right]v_{3}\right\} \]

so that

\[
\left[\tilde{u}v\right]_{1}=\frac{G_{1}-G_{3}}{G_{1}}u_{2}v_{3}+\frac{G_{2}-G_{1}}{G_{1}}u_{3}v_{2},\cdots\]

The dots denote three more relations obtained by cyclic permutations.
In particular,

\[
[\tilde{v}v]_{1}=\frac{G_{2}-G_{3}}{G_{1}}v_{2}v_{3}\]
 The geodesic equations become

\[
\frac{dv_{1}}{dt}+\frac{G_{3}-G_{2}}{G_{1}}v_{2}v_{3}=0,\cdots\]

Calculating as above gives the formula for curvature

\[
R(u,v)=\frac{(G_{2}-G_{3})^{2}+2G_{1}(G_{2}+G_{3})-3G_{1}^{2}}{4G_{1}}\left(u_{2}v_{3}-v_{2}u_{3}\right)^{2}+\cdots\]

In particular, if we choose $u=(0,\frac{1}{\sqrt{G_{2}}},0)$to be
a unit vector in the second principal direction and $v$ to be a unit
vector in the third direction the sectional curvature of the $23-$plane
is

\[
K_{23}=\frac{(G_{2}-G_{3})^{2}+2G_{1}(G_{2}+G_{3})-3G_{1}^{2}}{4G_{1}G_{2}G_{3}}.\]

The others are given by cyclic permutations.

\subsubsection{Stability of the Rigid Body}

Even for an anisotropic rigid body with $G_{1}<G_{2}<G_{3},$ the
rotations around the principal axes are time-independent: $(v_{1},0,0)$
for example is a solution of the Euler equations \cite{LandauLifshitz}.
Under small perturbations, the rotations around the principal axes
with the largest and smallest moment of inertia are stable. But a
rotation around the principal axis with the middle value of moment
of inertia is unstable: small perturbations grow exponentially.

The geometric pitcure allows us to generalize this analysis to time-dependent
solutions. The geodesic deviation equation shows that a small perturbation
along $u$ to a geodesic with tangent vector $v$ will grow exponentially
if $R(u,v)<0.$ Let us see under what conditions this is possible.
A change of variables from the principal moments of inertia to the
principal curvatures of the body will help us.

Recall that the principal moments of inertia are

\[
G_{1}=M_{2}+M_{3},\cdots\]

where $M_{1},M_{2},M_{3}$ are the (always positive) eigenvalues of
the moment matrix $M=\int\rho x\otimes xdx$.

Define their reciprocals \[
\mu_{1}=\frac{1}{M_{1}}=\left[\int\rho(x)x_{1}^{2}dx\right]^{-1},\cdots\]

In terms of these,

{\tiny \[
(G_{2}-G_{3})^{2}+2G_{1}(G_{2}+G_{3})-3G_{1}^{2}=4\left[M_{1}M_{2}+M_{1}M_{3}-M_{2}M_{3}\right].\]
 }{\tiny \par}

{\tiny \[
=\frac{4}{\mu_{1}\mu_{2}\mu_{3}}\left[\mu_{2}+\mu_{3}-\mu_{1}\right]\]
 }{\tiny \par}

\[
R(u,v)=\frac{\mu_{2}+\mu_{3}-\mu_{1}}{\mu_{1}(\mu_{2}+\mu_{3})}[u_{2}v_{3}-u_{3}v_{2}]^{2}+\cdots\]

Thus, if the three principals $\mu_{1},\mu_{2},\mu_{3}$ satisfy the
triangle inequalities,

\[
\mu_{1}+\mu_{2}>\mu_{3},\quad\mu_{2}+\mu_{3}>\mu_{1},\quad\mu_{3}+\mu_{1}>\mu_{2}\]

the curvature $R(u,v)$ will be positive for all pairs: a rigid body
that is not `too anisotropic' is geodesically stable. But these inequalities
can be violated for a `flat enough' shape. Let us look at an example.

\subsubsection*{Stability of a rotating cylinder}

If the body has an axis of symmetry (say the third axis) $\mu_{1}=\mu_{2}$
and

\[
R(u,v)=\frac{\mu_{3}}{\mu_{1}(\mu_{1}+\mu_{3})}\left\{ [u_{2}v_{3}-u_{3}v_{2}]^{2}+[u_{1}v_{3}-u_{3}v_{1}]^{2}\right\} \]

\[
+\frac{2\mu_{1}-\mu_{3}}{2\mu_{1}\mu_{3}}[u_{1}v_{2}-u_{2}v_{1}]^{2}\]

So an instability arises if

\[
\mu_{3}>2\mu_{1}.\]

Consider a rotating cylinder of uniform density. Its height is $h$
and its radius is $r.$ It is elementary that

\[
M_{1}=M_{2}=\frac{r^{2}}{4},\quad M_{3}=\frac{h^{2}}{12}.\]

\[
\mu_{1}=\mu_{2}=\frac{4}{r^{2}},\quad\mu_{3}=\frac{12}{h^{2}}.\]

Thus a cylinder with too small a height

\[
h\leq\sqrt{\frac{3}{2}}r\]

will be unstable. If it is rotating around one of its diameters, it
is unstable with respect to a change of axis towards one of the other
diameters. An example of such motion is a coin toss.

\section{The Geometry of Diffeomorphisms}

It is a remarkable fact that the set of Diffeomorphisms of a Riemannian
manifold is itself a Riemannian manifold. This \textquotedbl{}higher
Riemannian geometry\textquotedbl{} is the proper language of many
interesting physical systems such as ideal fluids. Such repetitions
of structures at a higher level happen quite often in mathematics:
the set of Riemannian metrics on a manifold is itself a Riemmannian
manifold; the set of complex structures is a complex manifold; the
set of Kähler structures is itself a Kähler manifold and so on.

\subsection{The Circle}

The simplest manifold is the circle. So the first example of a Diffeomorphism
group must be $Diff(S^{1}).$ The standard metric on the circle leads
to the $L^{2}-$metric on this group.

\[
G(u,v)=\int_{0}^{2\pi}u(x)v(x)dx.\]

The deformation tensor of a vector field on the circle is easily found
from

\[
G(\tilde{u}v,w)=G(uv'-vu',w)+G(v,uw'-wu')=\int\left\{ uv'w-vu'w+vuw'-vwu'\right\} dx.\]

to be

\[
\tilde{u}=-3u'.\]

Thus the covariant derivative on the Lie algebra is

\[
D_{u}v=\frac{1}{2}\left\{ [u,v]-\tilde{u}v-\tilde{v}u\right\} =u'v+2uv'.\]

The geodesic equation is the inviscid Burger's equation

\[
\frac{\partial v}{\partial t}+3v\frac{\partial v}{\partial x}=0.\]

The curvature form

\[
R(u,v)=\frac{1}{4}|[u,v]+\tilde{v}u-\tilde{u}v|^{2}+G(\tilde{u}v,\tilde{v}u)-G\left(\tilde{u}u,\tilde{v}v\right)\]

reduces to

\[
R(u,v)=|[u,v]|^{2}.\]

Thus the $L^{2}$-metric on $Diff(S^{1})$ has positive curvature.

\subsubsection*{A Technical Remark}

This contradicts the statement in \cite{Kambe} and in several other
places in the mathematics literature that the $L^{2}-$ metric on
the group of diffeomorphisms of $M$ has zero curvature if $(M,g)$
is itself flat. In particular, it is not possible to calculate the
curvature of the subgroup of incompressible diffeomorphisms by using
the Gauss-Codazzi formula as is often claimed. This is despite the
fact the final answer has an uncanny resemblance to this formula .
The confusion arises because there is a connection on $Diff(M)$ whose
curvature form is the average of the curvature form of $M$. But it
does not preserve the metric on the group manifold:

\[
\int g(\nabla_{u}v,w)dv_{g}+\int g(v,\nabla_{v}w)dv_{g}\neq0\]

where $dv_{g}$ is the Riemannian volume element on $M$. We take
a more direct route in the next section to avoid this pitfall.

\subsection{Incompressible Diffeomorphisms}

Given a scalar density on a manifold, the set of diffeormorphisms
that preserve it form a subgroup:

\[
\mathfrak{D}_{\rho}=\left\{ \phi\in\mathfrak{D}|\rho(x)=\det\partial\phi(x)\rho\left(\phi(x)\right)\right\} .\]

Its Lie algebra is the set of vector fields of zero divergence. Recall
that the divergence of a vector field is defined with respect to a
density as

\[
\mathrm{div}u=\frac{1}{\rho}\partial_{i}\left[\rho u^{i}\right].\]

From the identity

\[
\mathrm{div}[u,v]=u\left[\mathrm{div}v\right]-v\left[\mathrm{div}u\right]\]

it follows that incompressible (divergenceless) vector fields form
a sub-Lie algebra, which we will call $\mathcal{G}$.

Given a Riemannian metric $g$ on $M$ there is an inner product on
the space of vector fields on $M$:

\[
G(u,v)=\int g(u,v)\rho.\]

We will call this the $L^{2}-$inner product even though we are not
completing the space of vector fields with respoect to the induced
norm. Define now the \emph{deformation tensor} $\tilde{u}$ of a vector
field:

\[
G(\tilde{u}v,w)=\int\left\{ g([u,v],w)+g(v,[u,w])\right\} \rho\]

\[
=\int\left\{ g(\nabla_{u}v-\nabla_{v}u,w)+g(v,\nabla_{u}w-\nabla_{w}u)\right\} \rho\]

where $\nabla_{u}$ is the Riemannian covariant derivative on the
manifold $(M,g).$ Thus

\[
G(\tilde{u}v,w)=\int\left\{ \nabla_{u}\left[g(v,w)\right]\right\} \rho-\int v^{i}w^{j}\left[\nabla_{i}u_{j}+\nabla_{j}u_{i}\right]\rho\]

After an integration by parts the first term is zero, when $u$ has
zero divergence. It follows that

\[
[\tilde{u}v]_{j}=-\left[\nabla_{i}u_{j}+\nabla_{j}u_{i}\right]v^{j}+\nabla_{i}\phi(u,v)\]

where $\phi(u,v)$is to be chosen such that the lhs has zero divergence:

\[
\nabla^{i}\left\{ -\left[\nabla_{i}u_{j}+\nabla_{j}u_{i}\right]v^{j}+\nabla_{i}\phi(u,v)\right\} =0\]

Thus, absorbing $\nabla(g(u,v))$into the gradient term,

\[
[\tilde{u}v+\tilde{v}u]=-[\nabla_{v}u+\nabla_{u}v]+\nabla_{i}p(u,v)\]

\[
[D_{u}v]=\frac{1}{2}\left\{ [u,v]+\nabla_{v}u+\nabla_{u}v\right\} +\nabla p(u,v)\]

But

\[
[u,v]=\nabla_{u}v-\nabla_{v}u\]

so that

\[
[D_{u}v]=\nabla_{v}u+\nabla p(u,v),\quad\nabla^{2}p(u,v)+\mathrm{div}\nabla_{v}u=0.\]

In particular

\[
D_{v}v=\nabla_{v}v+\nabla p\]

which $p$ chosen such that the divergence is zero. So the geodesic
equation on the group of volume preserving diffeomorphisms is the
Euler equation, as promised.

The $L^{2}-$inner product allows us to split the space of vector
fields into the subspace of divergence-free vector fields (which is
also a subalgebra) and another subspace of gradients:

\[
u=u^{T}+\nabla\phi(u),\quad\mathrm{div}u^{T}=0,\quad\nabla^{2}\phi(u)=\mathrm{div}u.\]

This is an orthogonal decomposition.The Riemannian covariant derivative
in the group of volume preserving diffeomorphisms is just the transverse
projection of the covariant derivative on $(M,g)$:

\[
D_{u}v=\left[\nabla_{u}v\right]^{T}.\]

Thus we can split the $L^{2}-$inner product on vector fields into
transverse and longitudinal pieces

\[
G(u,v)=T(u,v)+S(u,v)\]

where

\[
T(u,v)=G(u^{T},v^{T}),\quad S(u,v)=G(\nabla\phi(u),\nabla\phi(v)).\]

In other words, even if $w$ is not of zero divergence

\[
G(D_{u}v,w)=T\left(\nabla_{u}v,w\right)\]

\subsection{Curvature of the Diffeomorphism Group}

\[
R(u,v)=G\left({D}_{[u,v]}u,v\right)-G\left({D}_{u}{D}_{v}u,v\right)+G\left({D}_{v}{D}_{u}u,v\right)\]

\[
=G\left({D}_{[u,v]}u,v\right)+G\left({D}_{v}u,D_{u}v\right)-G\left({D}_{u}u,D_{v}v\right)\]

\[
=G\left(\nabla_{[u,v]}u,v\right)+T\left(\nabla_{v}u,\nabla_{u}v\right)-T\left(\nabla_{u}u,\nabla_{v}v\right)\]

where we use the fact that $v$ has zero divergence.

Now,

\[
\nabla_{[u,v]}u=\]

\[
=r(u,v)u+\nabla_{u}\nabla_{v}u-\nabla_{v}\nabla_{u}u\]

\[
R(u,v)=\int r(u,v)\rho+G\left(\nabla_{u}\nabla_{v}u-\nabla_{v}\nabla_{u}u,v\right)+T\left(\nabla_{v}u,\nabla_{u}v\right)-T\left(\nabla_{u}u,\nabla_{v}v\right)\]

Set $w=\nabla_{v}u,$ which may not have zero divergence, even though
$u$ and $v$ are of zero divergence. Then,

\[
G\left(\nabla_{u}w,v\right)=\int g\left(\nabla_{u}w,v\right)\rho\]

\[
=\int\nabla_{u}\left[g(w,v)\right]\rho-\int g\left(w,\nabla_{u}v\right)\rho\]

\[
=-G(w,\nabla_{u}v)=-S(w,\nabla_{u}v)-T(w,\nabla_{u}v)\]

And similarly for $G(\nabla_{v}\nabla_{u}u,v).$ Thus

\[
R(u,v)=\bar{r}(u,v)+S(\nabla_{u}u,\nabla_{v}v)-S(\nabla_{v}u,\nabla_{u}v)\]

where

\[
\bar{r}(u,v)=\int r(u,v)\rho\]

is the curvature form of $(M,g)$ averaged by the density. It is not
difficult now to work out explicit answers for a flat torus and recover
Arnold's original results. We find the present form more useful as
well as more general.

\section{Conclusion }

Arnold's geometric explanation of the instabilities of an ideal fluid
raises an important physical question. Do the Euler equations describe
the time evolution of real fluids? Any theoretical description of
a physical system is an idealization in which small forces are ignored.
This is usually justified because small forces have small effects.

Near an unstable point this breaks down . Just think of a pendulum
balanced precariously on its head, at the unstable equilibrium point.
A small force in either direction can topple it in one direction or
the other . The final outcome is random, since no one can predict
with certainty such tiny perturbations.

What if a system is dynamically unstable at every point in its phase
space? This is the case with an ideal fluid. In this case no where
is it justified to ignore the unpredicatable small forces. Instead
of predicting these forces, we must model them as a stochastic process;
for example, as Gaussian white noise. The time evolution of the system
in then a stochastic differential equation.

What is the evolution of an unstable dynamical system under infinitesimally
small white noise random forces? In a later paper \cite{RajeevInPrep}
we will show that this can be reduced to a \emph{deterministic} dynamical
system, but with \emph{twice as many} degrees of freedom. The Euler
equations will then get replaced with equations for geodesics on the
tangent bundle of the diffeomorphism group.

\section{Acknowledgements}

I thank V. V. Sreedhar for organizing this series of lectures at the
Chennai Mathematical Institute; M. Gordina for an invitation to speak
at the University of Connecticutt; both of them and A. Agarwal, B.
Driver, L. Gross, A. Joseph, D. Karabali, G. Krishnaswami, G. Misiolek,
V. P. Nair, A. Polychronakos and T. Turgut for several discussions
of this and many related topics.

\end{document}